\documentstyle[twoside,psfig]{article}

\catcode`\@=11
\long\def\@makefntext#1{
\protect\noindent \hbox to 3.2pt {\hskip-.9pt  
$^{{\eightrm\@thefnmark}}$\hfil}#1\hfill}		

\def\@makefnmark{\hbox to 0pt{$^{\@thefnmark}$\hss}}	
	
\def\ps@myheadings{\let\@mkboth\@gobbletwo
\def\@oddhead{\hbox{}
\rightmark\hfil\eightrm\thepage}   
\def\@oddfoot{}\def\@evenhead{\eightrm\thepage\hfil
\leftmark\hbox{}}\def\@evenfoot{}
\def\sectionmark##1{}\def\subsectionmark##1{}}

\oddsidemargin=\evensidemargin
\newcounter{sectionc}\newcounter{subsectionc}\newcounter{subsubsectionc}
\renewcommand{\section}[1] {\vspace{12pt}\addtocounter{sectionc}{1} 
\setcounter{subsectionc}{0}\setcounter{subsubsectionc}{0}\noindent 
	{\tenbf\thesectionc. #1}\par\vspace{5pt}}
\renewcommand{\subsection}[1] {\vspace{12pt}\addtocounter{subsectionc}{1} 
	\setcounter{subsubsectionc}{0}\noindent 
	{\bf\thesectionc.\thesubsectionc. {\kern1pt \bfit #1}}\par\vspace{5pt}}
\renewcommand{\subsubsection}[1] {\vspace{12pt}\addtocounter{subsubsectionc}{1}
	\noindent{\tenrm\thesectionc.\thesubsectionc.\thesubsubsectionc.
	{\kern1pt \tenit #1}}\par\vspace{5pt}}
\newcommand{\nonumsection}[1] {\vspace{12pt}\noindent{\tenbf #1}
	\par\vspace{5pt}}

\newcounter{appendixc}
\newcounter{subappendixc}[appendixc]
\newcounter{subsubappendixc}[subappendixc]
\renewcommand{\thesubappendixc}{\Alph{appendixc}.\arabic{subappendixc}}
\renewcommand{\thesubsubappendixc}
	{\Alph{appendixc}.\arabic{subappendixc}.\arabic{subsubappendixc}}

\renewcommand{\appendix}[1] {\vspace{12pt}
        \refstepcounter{appendixc}
        \setcounter{figure}{0}
        \setcounter{table}{0}
        \setcounter{lemma}{0}
        \setcounter{theorem}{0}
        \setcounter{corollary}{0}
        \setcounter{definition}{0}
        \setcounter{equation}{0}
        \renewcommand{\thefigure}{\Alph{appendixc}.\arabic{figure}}
        \renewcommand{\thetable}{\Alph{appendixc}.\arabic{table}}
        \renewcommand{\theappendixc}{\Alph{appendixc}}
        \renewcommand{\thelemma}{\Alph{appendixc}.\arabic{lemma}}
        \renewcommand{\thetheorem}{\Alph{appendixc}.\arabic{theorem}}
        \renewcommand{\thedefinition}{\Alph{appendixc}.\arabic{definition}}
        \renewcommand{\thecorollary}{\Alph{appendixc}.\arabic{corollary}}
        \renewcommand{\theequation}{\Alph{appendixc}.\arabic{equation}}
        \noindent{\tenbf Appendix \theappendixc #1}\par\vspace{5pt}}
\newcommand{\subappendix}[1] {\vspace{12pt}
        \refstepcounter{subappendixc}
        \noindent{\bf Appendix \thesubappendixc. {\kern1pt \bfit #1}}
	\par\vspace{5pt}}
\newcommand{\subsubappendix}[1] {\vspace{12pt}
        \refstepcounter{subsubappendixc}
        \noindent{\rm Appendix \thesubsubappendixc. {\kern1pt \tenit #1}}
	\par\vspace{5pt}}

\newcommand{\textlineskip}{\baselineskip=13pt}
\newcommand{\smalllineskip}{\baselineskip=10pt}

\def\eightcirc{
\begin{picture}(0,0)
\put(4.4,1.8){\circle{6.5}}
\end{picture}}
\def\eightcopyright{\eightcirc\kern2.7pt\hbox{\eightrm c}} 

\newcommand{\copyrightheading}[1]
	{\vspace*{-2.5cm}\smalllineskip{\flushleft
	{\footnotesize Modern Physics Letters A, #1}\\
	{\footnotesize $\eightcopyright$\, World Scientific Publishing
	 Company}\\
	 }}

\newcommand{\publisher}[2]{{\begin{center}\footnotesize\smalllineskip 
	Received #1\\
	Revised #2
	\end{center}
	}}

\def\abstracts#1#2#3{{
	\centering{\begin{minipage}{4.5in}\footnotesize\baselineskip=10pt
	\parindent=0pt #1\par 
	\parindent=15pt #2\par
	\parindent=15pt #3
	\end{minipage}}\par}} 

\def\keywords#1{{
	\centering{\begin{minipage}{4.5in}\footnotesize\baselineskip=10pt
	{\footnotesize\it Keywords}\/: #1
	 \end{minipage}}\par}}

\renewenvironment{thebibliography}[1]
	{\frenchspacing
	 \ninerm\baselineskip=11pt
	 \begin{list}{\arabic{enumi}.}
        {\usecounter{enumi}\setlength{\parsep}{0pt}     
	 \setlength{\leftmargin 12.7pt}{\rightmargin 0pt} 
         \setlength{\itemsep}{0pt} \settowidth
	{\labelwidth}{#1.}\sloppy}}{\end{list}}

\newcounter{itemlistc}
\newcounter{romanlistc}
\newcounter{alphlistc}
\newcounter{arabiclistc}

\newcommand{\fcaption}[1]{
        \refstepcounter{figure}
        \setbox\@tempboxa = \hbox{\footnotesize Fig.~\thefigure. #1}
        \ifdim \wd\@tempboxa > 5in
           {\begin{center}
        \parbox{5in}{\footnotesize\smalllineskip Fig.~\thefigure. #1}
            \end{center}}
        \else
             {\begin{center}
             {\footnotesize Fig.~\thefigure. #1}
              \end{center}}
        \fi}

\newcommand{\tcaption}[1]{
        \refstepcounter{table}
        \setbox\@tempboxa = \hbox{\footnotesize Table~\thetable. #1}
        \ifdim \wd\@tempboxa > 5in
           {\begin{center}
        \parbox{5in}{\footnotesize\smalllineskip Table~\thetable. #1}
            \end{center}}
        \else
             {\begin{center}
             {\footnotesize Table~\thetable. #1}
              \end{center}}
        \fi}

\def\@citex[#1]#2{\if@filesw\immediate\write\@auxout
	{\string\citation{#2}}\fi
\def\@citea{}\@cite{\@for\@citeb:=#2\do
	{\@citea\def\@citea{,}\@ifundefined
	{b@\@citeb}{{\bf ?}\@warning
	{Citation `\@citeb' on page \thepage \space undefined}}
	{\csname b@\@citeb\endcsname}}}{#1}}

\newif\if@cghi
\def\cite{\@cghitrue\@ifnextchar [{\@tempswatrue
	\@citex}{\@tempswafalse\@citex[]}}
\def\citelow{\@cghifalse\@ifnextchar [{\@tempswatrue
	\@citex}{\@tempswafalse\@citex[]}}
\def\@cite#1#2{{$\null^{#1}$\if@tempswa\typeout
	{IJCGA warning: optional citation argument 
	ignored: `#2'} \fi}}

\def\pmb#1{\setbox0=\hbox{#1}
	\kern-.025em\copy0\kern-\wd0
	\kern.05em\copy0\kern-\wd0
	\kern-.025em\raise.0433em\box0}

\def\fnt#1#2{\footnotetext{\kern-.3em
	{$^{\mbox{\scriptsize #1}}$}{#2}}}

\def\fpage#1{\begingroup
\voffset=.3in
\thispagestyle{empty}\begin{table}[b]\centerline{\footnotesize #1}
	\end{table}\endgroup}

\def\runninghead#1#2{\pagestyle{myheadings}
\markboth{{\protect\footnotesize\it{\quad #1}}\hfill}
{\hfill{\protect\footnotesize\it{#2\quad}}}}
\font\tenrm=cmr10
\font\tenit=cmti10 
\font\tenbf=cmbx10
\font\bfit=cmbxti10 at 10pt
\font\ninerm=cmr9

\font\eightrm=cmr8

\def\qed{\hbox{${\vcenter{\vbox{			
   \hrule height 0.4pt\hbox{\vrule width 0.4pt height 6pt
   \kern5pt\vrule width 0.4pt}\hrule height 0.4pt}}}$}}

\begin{document}
\setlength{\textheight}{7.7truein}  

\runninghead{Cosmic Censorship$\ldots$}{Cosmic Censorship$\ldots$}

\normalsize\textlineskip
\thispagestyle{empty}
\setcounter{page}{1}

\copyrightheading{}			

\vspace*{0.88truein}

\fpage{1}
\centerline{\bf COSMIC CENSORSHIP: A CURRENT PERSPECTIVE}
\baselineskip=13pt
\vspace*{0.37truein}
\centerline{\footnotesize PANKAJ S.
JOSHI\footnote{e-mail: psj@tifr.res.in}}
\baselineskip=12pt
\centerline{\footnotesize\it Tata Institute of Fundamental Research}
\baselineskip=10pt
\centerline{\footnotesize\it Homi Bhabha Road, Mumbai 400 005} 
\centerline{\footnotesize\it India} 
\vspace*{10pt}


\publisher{(received date)}{(revised date)}

\vspace*{0.21truein}
\abstracts{End state of gravitational collapse and the related cosmic 
censorship conjecture continue to be amongst the most important
open problems in gravitation physics today. My purpose here is to bring out
several aspects related to gravitational collapse and censorship, 
which may help towards a better understanding of the issues involved. 
Possible physical constraints on gravitational collapse scenarios are
considered. It is 
concluded that the best hope for censorship lies in analyzing the 
genericity and stability properties of the currently known classes of 
collapse models which lead to the formation of naked singularities, rather 
than black holes, as the final state of collapse and which develop from 
a regular initial data.}{}{}

\vspace*{10pt}


\vspace*{1pt}\textlineskip	
\section{Introduction}	
\vspace*{-0.5pt}
\noindent
One of the most outstanding problems in the relativistic astrophysics 
and gravitation theory today is the final fate of a massive star, 
which enters the state of an endless gravitational collapse once it has 
exhausted its nuclear fuel. What will be the end state of such a 
continual collapse which is entirely dominated by the force of gravity? 
The conjecture that such a collapse, under physically realistic 
conditions, must end into the formation of a black hole is called the 
{\it cosmic censorship hypothesis}(CCH). Despite numerous attempts over the
past three decades, such a conjecture remains unproved and continues to
be a major unsolved problem, lying at the foundation of the theory
and applications in black hole physics.

Considering the failure of many attempts to establish the censorship 
conjecture, it would seem natural to arrive at the conclusion that what 
is really necessary here is to understand better and in a more extensive
manner the actual dynamical gravitational collapse process within 
the framework of general theory of relativity. Such efforts have taken 
place over the past decade or so, and the conclusion that is emerging 
is that the final fate of a continual collapse would be either a black 
hole (BH), or a naked singularity (NS), depending on the nature of the 
regular initial data, from which the collapse develops evolving from an
initial spacelike surface. For a discussion on various aspects of this  
problem and for some recent reviews, we refer to [1-4], and also
[5] for a detailed technical discussion on CCH).

We need to clarify here what we mean by occurrence of a naked
singularity developing as end point of collapse. At times, the 
non-existence of trapped surfaces till the formation of the singularity
in collapse is taken as the signature that the singularity is naked.
However, this need not be the case (see e.g. [1-2] for details). What
we mean by the development of a naked singularity in collapse is that
there exist families of future directed non-spacelike curves, which
in the past terminate at the singularity. No such families exist
originating from the singularity when the end product of collapse is a
black hole. In the case of a 
black hole forming, the resultant spacetime singularity will be hidden
inside an event horizon of gravity, remaining unseen by external
observers. On the other hand, if the collapse ends in a naked or
visible singularity, there is a causal connection between the region of
singularity and faraway observers, thus enabling in principle a
communication from the super dense regions close to singularity to 
faraway observers.

My purpose here is to examine and discuss several  
issues involved here which clarify the implications of some of the work in 
this area 
done so far, and to see where we stand as far as the censorship 
hypothesis is concerned.  We also point 
out, by drawing on examples from the models analyzed so far, that it 
is not possible to rule out the occurrence of naked singularities in 
collapse just based on simple physical reasonings, and that the problem 
is deeper. The point of view that emerges is as far as
the occurrence is concerned, both black holes and 
naked singularities appear to be basic properties consequent from the 
dynamics of Einstein equations, emerging in a natural manner and as a 
logical consequence of the general theory of relativity. 
However, the crucial issue is that of genericity and stability of
naked singularities. It would appear that the real hope for censorship 
lies in investigating in detail the stability properties of the 
collapse models which develop into naked singularity.

In the next section, we consider several physical conditions 
that one may like to impose on a physically realistic gravitational
collapse scenario, and we point out the status of CCH {\it vis-a-vis} such
constraints. In Section 3 we discuss three further possibilities, which
I think offer a more serious avenue as far as the CCH is concerned. 
In the concluding section, it is emphasized that it might help
to try to get a better insight into the phenomenon of naked singularity 
formation, that is, we may try to understand better why actually naked
singularities develop in gravitational collapse.

\section{Physical Constraints on Gravitational Collapse}
\noindent
As noted above, in any general discussion on CCH, it is stated that
naked singularities do not form in collapse under {\it physically
realistic
conditions}. However, the precise physical conditions under which the 
censorship is supposed to be holding are usually not specified. The 
advantage then is that even if a certain set of physical conditions 
did not work towards proving CCH, one still has the option to try out 
another set of physical constraints to continue further efforts.

If by such a procedure we are to prove censorship in some suitable 
form, we have to eventually arrive at a proper and appropriate mathematical 
formulation of the censorship conjecture we want to establish. As of 
today, a suitable formulation of CCH is a major problem in itself. Many
natural looking physical conditions have been proposed and tried out,
these being indicated as a remedy to rule out naked singularities. This 
is with the hope to arrive ultimately at a suitable formulation of CCH.

We examine below several such physical constraints on a realistic 
gravitational collapse scenario, and the implications they have towards 
determining the final fate of collapse. In particular, the motivation is
to rule out NS as final state by imposing such conditions. It turns out
that one still does not succeed in ruling out NS with the help of such 
conditions considered so far. But the advantage of such an analysis 
is, firstly, it clarifies the situation as to what such conditions 
can possibly achieve. Secondly, 
it serves as a pointer to some thing deeper we should look for if we are 
to establish CCH. Finally, this also implies that in fact NS
do develop in wide classes of gravitational collapse scenarios under 
realistic 
physical conditions. We also try to get an insight here why many of 
such conditions 
have not worked, or are unlikely to work towards establishing CCH, 
and why we must explore further more subtle alternatives. Eventually,
we point out that the hope for CCH appears to lie in a detailed
genericity and stability analysis only of the available collapse models 
which result in a naked singularity formation.

\subsection{A suitable energy condition must be obeyed}
\noindent
This is one of the basic conditions assumed in the classical gravity
description, and should be satisfied by the matter fields constituting 
the star at least till the collapse has proceeded to such an advanced 
stage so as to enter a phase governed by quantum gravity, that is, till
the classical description starts breaking down in one way or the other.

In fact, if one allowed for completely arbitrary matter fields, it will
be quite easy to produce naked singularities. For example, start with a 
geometry allowing families of future directed non-spacelike geodesics,
which are future endless, but terminate in the past at the singularity.
Then {\it define} the matter fields to be given by,

\begin{equation}
T_{ij} \equiv {1\over 8\pi} G_{ij}. \label{one}
\end{equation}

Hence, it is obvious that one must consider scenarios where matter
fields do satisfy reasonable physical conditions. One would hope that
a suitable energy condition would be one of these, as all observed
classical fields do obey such a condition. A further motivation would be 
energy conditions have been used extensively in the singularity theorems
in general relativity, which predict the existence of singularities in
gravitational collapse and cosmology. 

One would like to see if CCH is obeyed once we have assumed matter 
fields to satisfy suitable energy conditions. It turns out, however,  
that there are several classes of collapse models, where in fact collapsing 
matter does satisfy a proper energy condition, but the collapse does 
end in a naked singularity (henceforth, when additional references are 
not given then [1-4] and [5] may be looked up for details).

Actually, there are classes of collapse models where satisfying 
the energy condition appears
to be aiding the naked singularity formation as the end state of collapse, 
in turn making it physically more interesting and serious. An
example of this is
the spherically symmetric self-similar collapse of a perfect fluid. The
general form of metric is,
\begin{equation}
ds^2 = -e^{2\nu(r,t)} dt^2 + e^{2\psi(r,t)} dr^2 + r^2S^2(r,t)(d\theta^2 + 
sin^2\theta d\phi^2) \label{}
\end{equation}
where the metric functions depend on $X=t/r$ due to self-similarity.
One can work out the outgoing null geodesics from the naked
singularity, which turn out to be related to the density and pressure
distributions in the spacetime via the Einstein equations. These are
then given by,
\begin{equation}
r = D(X-X_0)^{2\over {H_{0}-2}}\label{}
\end{equation}
Here $H_0$ is the limiting value of the quantity $H= (\eta + p)e^{2\psi}$,
$\eta$ and $p$ correspond to density and pressure, and $D>0$ is constant
of integration.
The weak energy condition is then equivalent to the statement that
$H_0>0$, which in turn ensures, from the above geodesics equation, that
{\it families} of null geodesics, as opposed to single isolated curves,
come out from the naked singularity at $t=0,r=0$, which is a node in
the $(t,r)$ plane.

It has to be noted that the Einstein equations as such do not 
require or impose an energy condition on matter distributions. It is
a criterion motivated on purely physical grounds. This then suggests
another possibility, namely that if some how in the later stages of
collapse the energy conditions are violated through whatever agency,
there may then be a hope to preserve CCH. What we mean is, for example 
in the models discussed above, in the equation for null geodesics if 
we violate energy conditions that would be corresponding to a 
negative value of $H_0$, and then there are no outgoing null geodesics 
from the singularity and CCH is essentially preserved.
It may be noted here that for a quantum fluid or a quantum field,
which may have important role to play in the very late stages of collapse,
the weak energy condition need not be obeyed, and it may be worth 
speculating if this some how may help save cosmic censorship.

\subsection{The collapse must develop from regular initial data}
\noindent
This is one of the most important physical constraints necessary 
for any possible version of CCH. If we are to model realistic collapse 
scenarios of matter clouds such as gravitationally collapsing massive 
stars, then the densities, pressures
and other physical quantities must be finite and regular at the initial
surface from which the collapse develops. That is, the initial 
spacelike surface should not admit any density or curvature singularities 
in the initial data so as to represent collapse from regular matter
distribution.

Generally, this is ensured by imposing the usual differentiability
conditions on the functions involved, together with requirements of 
finiteness and regularity. It is known by now that regular distributions 
of initial densities and pressures (for example, finite and suitably
differentiable on the initial surface) do give rise to both naked 
singularities and black holes, depending on the nature of the regular 
initial data from which the collapse evolves (see e.g. [4] for more
details). It turns out that given such an initial data, there are still
sufficient number of free functions available to choose in Einstein equations,
subject to weak energy condition and a suitable matching to the exterior
of the collapsing cloud, so that the evolution can end in either of
the BH/NS outcomes as desired.

At times, more stringent requirements are imposed on the initial data, 
e.g. asking for a complete smoothness of densities and pressures. Usually,
there are two motivations for this. One could be the requirements while
doing numerical evolutions, where smoothness (which is the same as 
demanding the analyticity of these functions) simplifies the analysis
considerably.
At other times, it is argued that astrophysically reasonable initial data
must be analytic. In the case of collapse of a dust cloud, this amounts to
demanding analyticity of the density function. The initial density
$\rho(r)$ then must contain no odd powers in $r$, and we have,
\begin{equation}
\rho(r) = \rho_0 + \rho_2 r^2 + \rho_4 r^4 + ...
\end{equation}
at the initial surface $t=t_i$, which gives an analytic density
profile. It is known, however, that even in the case of smooth density 
profiles with only even terms non-vanishing, the marginally bound 
dust evolution can end in a naked singularity (for example, when 
$\rho_2\ne0$), which is gravitationally strong, i.e. 
sufficiently fast divergence of curvatures does take place in the 
limit of approach to the singularity.

\subsection{Singularities from realistic collapse must be
gravitationally strong}
\noindent
This has been one of the most useful physical requirements, which
was explored rather thoroughly in order to develop a formulation for CCH.
The idea has been that any singularity that will develop from a realistic 
collapse has got to be physically serious in various aspects, including 
powerful divergences in all important physical quantities such as 
densities, pressures and curvatures etc, at least at the classical level. 
A typical condition for the singularity to be gravitationally strong is,
in addition to the divergences such as above, the gravitational tidal
forces must diverge and all physical volumes are crushed to zero size
in the limit of approach to the NS. A sufficient condition for this to 
happen is,
\begin{equation}
R_{ij}V^iV^j\propto {1\over k^2}
\end{equation}
where $k$ is the affine parameter along the non-spacelike geodesics
coming out from the singularity, with $k=0$ at the NS, and $V^i$ is 
the tangent vector to these curves emanating from NS.
Another criterion for strength was given recently by Nolan$^6$.

The singularity developing within the black hole formed out of the
standard dust cloud collapse as investigated by Oppenheimer and Snyder
is gravitationally strong 
in this sense. Now, if one could establish that whenever naked 
singularities formed in gravitational collapse, they are always 
gravitationally weak, in the sense of important divergences such as above 
not being present in the limit of approach to the singularity, then
such singularities could be removable from the spacetime, and one may 
be able to extend the spacetime through the same$^7$. Such removable
naked singularities should no longer be regarded as physically genuine,
and one has then established CCH in some form such as NS could develop
in gravitational collapse, however, they would be always gravitationally
weak and removable.

This possibility has been investigated thoroughly, and it is known 
now that gravitationally powerfully strong naked singularities actually
do result from collapse from regular initial data (including smooth
analytic density profiles), for several reasonable forms of matter such as
dust, perfect fluids, Vaidya radiation collapse, and several other forms
of matter satisfying suitable energy conditions. At such naked
singularities, the densities, curvature scalars such as the Kretschmann
scalar and gravitational tidal forces diverge most powerfully
as characterized above, which is as powerfully strong as the divergences 
observed at physical singularities such as the big-bang in cosmology.

\subsection{The matter fields must be sufficiently general}
\noindent
If NS formed in the collapse for certain special forms of matter only,
such as dust or collapsing radiations, that would not be much of interest.
For example, the role of pressures cannot be underestimated in 
realistic collapse and so one would like to know if matter with pressures
will necessarily give rise to a black hole only on undergoing gravitational
collapse. If such was the case, one could then rule out matter 
fields giving rise to NS as special or unphysical towards formulating 
CCH, even if they satisfied an energy condition or the collapse developed 
from regular initial data.

It is now known, however, that naked
singularities are not special to any particular form of matter field.
One can study the collapse for a general form of matter, the
so called type I matter fields (all the known physical forms of matter, 
such as dust, perfect fluids, massless scalar fields etc are included
in this class) subject to weak energy condition. The result is, given an
arbitrary but regular distribution of matter on the initial surface, there
are always evolutions available from this initial data which would either 
result in a black hole or naked singularity, depending on the allowed 
choice of free functions available from Einstein equations. More
specifically, in spherically symmetric collapse with a type I general 
matter field, given the distribution of density and the radial and tangential
pressure profiles on the initial surface, from which the collapse develops, 
one can then choose the free function describing the velocities of 
the in-falling shells in such a manner so as to have a black hole or a 
naked singularity as the final end product, depending on this choice
(see [4] and references therein for details).

\subsection{The collapsing cloud must obey a realistic equation of state}
\noindent
It is conjectured at times that even though naked singularities may 
develop for general matter fields, they must go away once a physically 
reasonable and realistic equation of state is chosen for the 
collapsing cloud.

This is a very difficult argument to formulate as it turns out. Firstly,
naked singularities {\it do} form in collapse of several well-known
equations of state, such as dust, perfect fluids, or in-flowing radiation
shells. Secondly, it is extremely difficult to make any guesses as to
what might be the state of matter, or the realistic equation of state 
within a collapsing body such as a massive star which is in its advanced
stages of collapse. Thirdly, the collapsing cloud may not have a single
equation of state, which might actually be changing as the collapse evolves.
There have been speculations, for example, that strange quark matter
may be a good approximation to the collapsing star in its final stages$^8$,
and the collapse was then examined in a Vaidya geometry, which again 
results in BH/NS phases as usual. In other words, such a choice of equation
of state also does not remove naked singularities. At the other extreme,
there are also arguments such as those given by Hagerdorn and Penrose,
that the equation of state, in the very final stages of collapse much 
closely approximate that of dust$^9$. In other words, at higher and higher
densities, matter may behave more and more like dust. The point is, if
pressures are not negative then they also may contribute positively to
collapse just to add to the dust effect, and may not alter the conclusions
arrived at in the dust case. In such a case, the dust collapse situation, 
which has been investigated rather thoroughly, would imply that both BH/NS 
phases would clearly develop in gravitational collapse, depending on the
initial density and velocity distributions.

The point is, while there are several widely used and familiar 
equations of state available which result in the formation of naked 
singularities as final fate of collapse, there is still not a single 
equation of state available so far which definitely ensures that the
end product will be necessarily a black hole only. Under the situation, 
one cannot help hazarding a guess that the crux of the matter may not 
lie in the equation of state or the form of matter collapsing, and 
over-emphasizing that particular option might amount eventually to 
barking under a wrong tree as far as the search for CCH is concerned, 
given the severe uncertainties on the state of the matter in the
very late stages of collapse, as described above.

\subsection{All radiations from naked singularity must be infinitely 
red-shifted}
\noindent
In certain sub cases of dust collapse resulting in naked singularity, 
it is seen that the red shift along the null geodesics coming out from
the NS diverges in the limit of approach to NS. This has given rise to 
the possibility that even if NS forms in collapse, no energy could 
escape from the same. In that sense, NS may be invisible to external
observers for all practical purposes. Of course, it has to be noted that
even if true, this does not save CCH in the actual sense, because
after all and basically CCH is about the question of principle in the 
general theory of relativity, namely whether singularities forming 
in gravitational collapse are causally connected to an external 
observer or not via nonspacelike trajectories.

In any case, it may be good to explore such a possibility,
because it may give some information on the structure of NS at least in 
certain special models, and if true generally, then it will provide some 
kind of a physical formulation for CCH. However, in my own perception,
it will be extremely difficult to establish in general that no energy
can come out from NS. There could be several reasons for this. Firstly, 
it may be quite tricky to apply the conventional definition of red shift, 
which corresponds to a regular source and observer, to emissions from 
a naked singularity. Secondly, even if there was no escape of energy 
along null geodesics, the possibility of mass emission via timelike or
non-spacelike non-geodetic families of paths coming out from the naked
singularity remains open. In the case of such a violent event being
visible, particles escaping with ultra relativistic velocities cannot be 
ruled out from the {\it neighborhood of NS}.

Apart from such technical difficulties, it is also to be noted that
the classical possibilities such as above regarding the probable light
or particle emission, or otherwise, from a naked singularity may not 
perhaps offer a serious physical alternative eventually 
one way or the other. The reason is, in all physical situations, the
classical general relativity should break down once the densities and   
curvatures are sufficiently high so that quantum or quantum
gravity effects should become important in the process of an endless 
collapse. Such quantum effects would come into play much before the 
actual formation of the classical naked singularity, which itself 
may possibly be smeared out by quantum gravity. The key question then is 
that of the possible visibility, or otherwise, of these extreme 
strong gravity regions, which develop in any case, in the vicinity of 
the classical naked singularity. It is then the causal structure, 
that is, the communicability or otherwise, of these extreme strong 
gravity regions that would make the essential difference as far as 
the physical consequences of a naked singularity are concerned, 
rather than aspects such as classical red shift$^{10}$.

\section{Other Alternatives}
\noindent
As we noted above, none of the physical conditions, such as those
discussed above, are quite able to effectively rule out naked singularities,
which in turn may lead us to some possible formulation of CCH, either
physical or a mathematical one. With 
each of the constraints such as above, we have counter-examples which 
obey such a physical constrain but produce NS as end state of dynamical 
collapse.

This brings us to three further possibilities which are under active
current investigation as of today towards a possible formulation of CCH,
and which I think may offer a better hope for CCH.
We now briefly discuss these below.

\subsection{Will quantum gravity remove naked singularities?}
\noindent
It is sometimes argued that after all the occurrence of singularities
is a classical phenomena, and that whether they are naked or covered
should not be relevant - quantum gravity will any way remove them all. 
But this is missing the real issue it would appear. 
It is possible that in a suitable quantum gravity theory the singularities 
will be smeared out (though this has been not realized so far, and 
also there are indications that in quantum gravity also the singularities
may not go away). However, in any case, the real issue is whether 
the extreme strong gravity regions formed due to gravitational collapse 
are visible to faraway observers or not. Because collapse certainly 
proceeds classically till the quantum gravity starts governing the 
situation at the scales of the order of Planck length, that is till the 
extreme gravity configurations have developed due to collapse. 
It is the visibility or otherwise of such regions that is under discussion.

The point is, classical gravity implies necessarily existence of 
strong gravity regions, where both classical and quantum gravity come 
into their own. In fact, as pointed out by Wald [2], if naked singularities
do develop in gravitational collapse, then in a literal sense we  
come face-to-face with the laws of quantum gravity whenever
such an event occurs in the universe. Then collapse phenomena 
has the potential to provide us with a possibility of actually testing 
the laws of quantum gravity.

In the case of a black hole developing in the collapse of a finite 
sized object such as a massive star, such strong gravity regions have
got to be necessarily hidden behind an event horizon of gravity, 
which would be well before the physical conditions became extreme. 
Then the quantum effects, even if they caused qualitative changes closer 
to singularity, will be of no physical consequences. This is because no 
causal communications are then allowed from such regions. 
On the other hand, if the causal structure were that of a NS, 
communications from such a quantum gravity dominated extreme curvature 
ball would be visible in principle, either directly or via secondary 
effects such as shocks produced in the surrounding medium.

\subsection{Should one consider all naked singularities produced by matter
fields to be unphysical?}
\noindent
There has been a suggestion$^{11}$ that all naked singularities, whenever
they are produced by matter fields such as dust, perfect fluids etc
should be rejected as being only `matter singularities', which should
have nothing to do with pure gravity. From such a perspective, the NS 
caused by massless scalar fields will be of course worrisome, which is 
included in the type I matter fields discussed above. While realistic stars 
are not made up of matter fields such as massless scalar field, may be
in the very final stages of collapse such matter forms may have important 
role to play in some manner.

It must be admitted, however, that not all will be comfortable with 
rejecting outright the logical consequences of collapse studies 
involving matter forms such as dust, perfect fluids, and such other fields.
After all, the classic gravitational collapse scenario, really at the
foundation of black hole physics and its chief motivator, is the 
homogeneous dust collapse model, as studied by Oppenheimer and 
Snyder$^{12}$.
Now, in the same models, when one puts in a density perturbation at the
center, then a naked singularity results, rather than a black hole. The
structure of the event and apparent horizons then change drastically so as
to expose the singularity to an external observer. Now, all realistic
stars will have a higher density at the center, falling off as some
rate as one moves away from the center. In that sense, one may want to 
regard the NS developing due to this density gradient at least as
physical as the black hole. After all, general relativists have worked
with dust and perfect fluids for several decades, and could be quite 
comfortable with the logical outcomes available within those collapse 
scenarios$^{13}$. Again, if one considers arguments such as those given
above in Sec. 2.5 in favour of equations of state such as dust in the
final phases of collapse, one may like to take the outcomes of such
a collapse physically more seriously.

\subsection{Are naked singularities stable and generic?}
\noindent
In my own opinion, this is the key issue on which any possible
future formulation and proof of the CCH would crucially depend. Even if
naked singularities do develop in collapse models, if they were not
generic and stable in some suitably well defined sense, that would make 
a good case for CCH. For example, most of the current classes of NS 
are within the framework of spherically symmetric collapse. While there
are some indications that NS do develop in non-spherical collapse as 
well$^{14}$, as such non-spherical collapse remains a largely uncharted
territory and it would be essential to examine it rather thoroughly.
In this connection, it is also to be noted that naked
singularities formed in the collapse of matter with positive energy are  
always `massless', in a sense obvious for spherical collapse, which
is yet to be made precise generically. Of course, even in that case,
one still needs to worry about the extreme high densities and curvatures
in that region that is visible as opposed to the BH case.

The key question one may then want to resolve here is, while we know that
physically reasonable initial data do give rise to naked singularities,
will the initial data subspace, which gives rise to NS as end state of
collapse have a zero measure in a suitable sense? As is well-known,
however, the stability theory in general relativity is a rather
complicated area, because there is no well-defined formulation or 
criteria to test stability. Before one could test CCH, a satisfactory
formulation for stability criterion has to be arrived at 
within the framework of general relativity. Also, the issue of what
is a suitable measure in the initial data space can be a complicated  
one. Only after making some reasonable progress here one could then 
start testing these questions for NS formation. While discussing
stability and genericity, one has also to be careful on the criterion 
one used to test the same, because sometimes a criterion can be used 
which makes black holes also unstable while trying to show the 
instability of naked singularities.

In the absence of such well-defined criteria against which to test
the available NS models, various attempts have been made to examine
if NS would be stable to some kind of perturbations. These include
perturbing the density profiles to include pressures, trying to see
how the density gradients at various levels affect global versus the
local visibility of the naked singularity, imposing symmetry conditions
such as self-similarity, and then to see how the conclusions change 
on relaxing the self-similarity condition, study of how certain
perturbations grow in the limit of approach to the Cauchy horizon, which
is the first ray coming out of the naked singularity, and such others.
While these attempts do not provide any definitive conclusions regarding
the stability or otherwise of NS, they surely provide a good insight
into the phenomena of BH/NH phases to tell us what is possible 
in gravitational collapse.

On the other hand, given the complexity of the Einstein field 
equations, if a phenomenon occurs so widely in spherical symmetry, it 
is not unlikely at all that the same would be repeated in more 
general situations as well. In fact, before the advent of
well-known singularity theorems in general relativity, it was widely
believed that the singularities found in more 
symmetric situations such as the Schwarzschild or Friedmann-Robertson-Walker
cosmological models will go away once we go to general enough spacetimes.
As is well-known, the singularity theorems then established that spacetime
singularities occur in rather general spacetime settings without symmetry
assumptions, and under a broad set of physical conditions. Thus, the
singularities which manifested earlier in symmetric situations were
actually indicative of a deeper phenomena. Such a possibility cannot again 
be
ruled out in the case of occurrence of naked singularities as well.

\section{Conclusions}
\noindent
In the above, we clarified the basic philosophy and motivation for 
cosmic censorship and the crucial role it plays in black hole physics.
We then outlined some of the approaches that have 
been tried out so far to formulate or prove the same. It turns out that 
none of the physical constraints or natural looking physical conditions 
are really able to ensure the validity of CCH. In fact, one tends to
conclude that naked singularities can actually develop in physically
realistic gravitational collapse situations.

It then follows that more radical options, such as those listed in the 
previous section, must be tried out if CCH is to be preserved.
We discussed these above, and it would appear that only one of them, 
namely that involving the stability and genericity of naked singularities 
can be a potentially promising candidate as far as any possible proof 
of CCH is concerned.

While one tries to work towards CCH along one of these or other paths, 
it is in fact important and quite interesting to really try to understand 
why do naked singularities 
actually develop in gravitational collapse. As we pointed out above,
several important physical constraints on collapsing clouds do not
appear to work towards helping CCH. It then becomes an intriguing  
question as to what is the physical agency that is possibly causing
NS in collapse in a rather natural manner within the framework of 
general relativity? Some work has been done recently in that 
direction$^{15}$, and it turns out that while gravitational collapse
proceeds, the shearing effects within the cloud could play a basic role 
to delay the formation of trapped surfaces and the apparent horizon in 
a natural manner. This in turn exposes the singularity to outside 
observers, depending on the rate of growth of shear in the limit of 
approach to the center. When looked at from such a perspective, one 
may even think that both black holes and naked singularities are rather 
natural consequences of gravitational collapse in classical general
relativity. Perhaps one can learn a lot on gravitational collapse 
by examining such physical processes that could be responsible for 
creation of naked singularity as against a black hole. 
It is asked many times that how could there be any 
other outcome other than a black hole possible as end state of 
collapse, 
when gravity is getting stronger and stronger. Delayed 
formation of trapped surfaces is then the answer, and it is the physical
processes like shear associated with the collapsing cloud that can
achieve this in a natural manner to resolve the `mystery' of naked
singularity formation in gravitational collapse.

The point is, trapped surface formation is intimately connected 
with the question of singularities. In singularity theorems, it is the 
crucial assumption. For CCH, the critical question is the epoch of its 
formation. An intuitive characterization of CCH would be then trapped 
surface 
formation precedes the singularity formation. The event of formation of 
trapped surface depends upon the dynamical properties 
of matter and does not depend so much on the equation of state, which 
refers to the general character of the matter. It rather depends upon much 
finer and detailed properties of matter distributions. 
In the light of such a reality, it is likely that it may 
not even be possible to charcterize CCH in a general form. May be
it can only be studied case by case for it depends on finer
structure and dynamics of the collapsing matter.
The main question then is what forms first, singularity or apparent 
horizon (trapped surface)? For NS, it should be singularity first and for 
BH it should be trapped surface first. Close to such events, matter would 
be in super dense state, which could have very unfamiliar behaviour and 
quantum effects could become dominant. This is what would perhaps drive 
future investigation in this area. Another important aspect is 
that of energy      
carried out by null rays emanating from the singularity. This is very        
important from the practical point of view, because singularity could in 
principle be naked yet harmless.

It is then possible that the cosmic censorship conjecture does not
hold classically, but may hold quantum mechanically$^{16}$, in some
sense yet to be figured out. What may be possible then is for a star
going into the final state of a naked singularity configuration, the
quantum gravity induced particle creation may take over to create a
burst like emission of energy, thus clearing up the 
naked singularity.

\nonumsection{Acknowledgments}
\noindent
Many of the contents above have been inspired by comments and issues that
came up at the IGQR meeting. I have greatly enjoyed the fruitful discussions 
with several friends including Dharam Ahluwalia, Jeeva 
Anandan, Sukratu Barve, Naresh Dadhich, Shyam Date, Hari Dass, Viquar 
Hussain, Sailo Mukherjee, Daniel Sudarsky, T. Padmanabhan, C. Unnikrishanan, 
and Matt Visser. I thank also Werner Israel and J. V. Narlikar for
their comments.

\nonumsection{References}

\end{document}